# Stellar Mass Black Holes and Ultraluminous X-Ray Sources

Rob Fender[1*] and Tomaso Belloni[2]

[1]Physics and Astronomy, University of South- ampton, Southampton SO17 1BJ, UK.

[2]Istituto Nazionale di Astrofisica–Osservatorio Astronomico di Brera, Via Emilio Bianchi 46, I-23807 Merate (LC), Italy.

* E-mail: r.fender@soton.ac.uk

*We review the likely population, observational properties, and broad implications of stellar-mass black holes and ultraluminous x-ray sources. We focus on the clear empirical rules connecting accretion and outflow that have been established for stellar-mass black holes in binary systems in the past decade and a half. These patterns of behavior are probably the keys that will allow us to understand black hole feedback on the largest scales over cosmological time scales.*

Stellar-mass black holes are the endpoints of the evolution of the most massive stars. The collapse of an iron core of >3 solar masses ($M_\odot$) cannot be stopped by either electron or neutron degeneracy pressure (which would otherwise result in a white dwarf, or neutron star, respectively). Within the framework of classical general relativity (GR), the core collapses to a singularity that is cloaked in an event horizon before it can be viewed. Like a giant elementary particle, the resulting black hole is then entirely described by three parameters: mass, spin, and charge (1). Because galaxies are old—the Milky Way is at least 13 billion years old—and the most massive stars evolve quickly (within millions of years or less), there are likely to be a large number of such stellar-mass black holes in our galaxy alone. Shapiro and Teukolsky (2) calculated that there were likely to be as many as 108 stellar mass black holes in our galaxy, under the assumption that all stars of initial mass >10 times that of the Sun met this fate.

The strongest evidence for the existence of this population of stellar-mass black holes comes from observations of x-ray binary systems (XRBs). In XRBs, matter is accreted (gravitationally captured into/onto the accretor), releasing large amounts of gravitational potential energy in the process. The efficiency of this process in releasing the gravitational potential energy is determined by the ratio of mass to radius of the accretor. For neutron stars, more than 10% of the rest mass energy can be released—a process more efficient at energy release than nuclear fusion (3). For black holes, the efficiency can be even higher, but the presence of an event horizon—from within which no signals can ever be observed in the out- side universe—means that this accretion power may be lost.

In some of these systems, dynamical measurements of the orbit indicate massive (>3 $M_\odot$) accretors that, independently, show no evidence for any emission from a solid surface. The first such candidate black hole x-ray binary (BHXRB) system detected was Cygnus X-1, which led to a bet between Kip Thorne and Stephen Hawking as to the nature of the accreting object (although both were moderately certain it was a black hole, Hawking wanted an insurance policy). In 1990, Hawking conceded the bet, accepting that the source contained a black hole. Since then, astronomers have discovered many hundreds of x-ray binaries within the Milky Way and beyond, several tens of which are good candidate BHXRBs (4).

Over the past decade, repeating empirical patterns connecting the x-ray, radio, and infrared emission from these objects have been found and used to connect these observations to physical components of the accretion flow (Fig. 1). It is likely that some of these empirical patterns of behavior also apply to accreting supermassive ($10^5$ to $10^9$ $M_\odot$) black holes in the centers of some galaxies, and that from studying BHXRBs on humanly accessible time scales, we may be learning about the forces that shaped the growth of galaxies over the lifetime of the universe. Between the stellar-mass black holes and the supermassive, there could be a population of intermediate-mass black holes (IMBHs), with masses in the range of $10^2$ to $10^5$ $M_\odot$. These may be related to the ultraluminous x-ray sources (ULXs), very luminous x-ray sources that have been observed in external galaxies. However, the problem of the nature of these sources is still unsettled, and alternative options involving stellar mass black holes are still open.

**Black Hole X-Ray Binaries**
There are several different approaches to classifying BHXRBs and their behavior, each of which can lead to different physical insights. One important approach is to look at the orbital parameters, and the most important of these is the mass of the donor star because it relates to the age of the binary. High-mass x-ray binaries have OB-type (5) massive donors and are young systems, typically with ages less than a million years or so. They are clustered close to the midplane of the Galactic disc and associated with star-forming regions. Cygnus X-1 is one of only a small number of high-mass BHXRB systems.

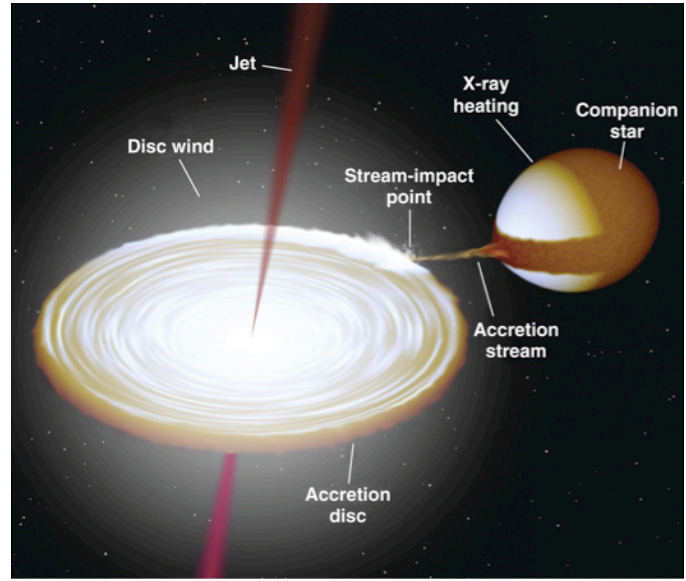

Fig. 1. An artist's impression of a low-mass BHXRB. The major components of the binary, accretion flow, and outflows are indicated. The inclination and relative masses of the binary components are based on estimates for the system GX 339-4, a key source in our understanding of black hole accretion and the source of the data presented in Fig. 2. This figure was produced by Rob Hynes.

Most BHXRBs are, however, low-mass x-ray binaries, which are almost certainly much older than the high-mass systems and as a result have a larger scale-height distribution in the galaxy and are hard to associate with their birth sites. In these systems, the companion donor star is of lower mass (typically less than 1 $M_\odot$) than the black hole. We now know of a sufficiently high number of accreting x-ray binaries to be able to study their galactic population, which we will not cover here. Most of these low-mass BHXRBs undergo transient outbursts, in that they typically spend most of their time in faint quiescent states before going into bright outbursts that approach the Eddington (6) limit and last from months to years. It is the detailed study of these outbursts, when the mass accretion rate onto the central black hole can change by orders of magnitude in just days, that has allowed us to make dramatic strides in our overall understanding of black hole accretion.

The outbursts of low-mass systems are likely to be triggered by a hydrogen ionization instability in the accretion disc while the mass transfer rate from the donor star remains steady. Initially, the mass transfer rate is greater than the rate of accretion onto the central black hole. The disc, initially neutral, slowly rises in temperature as the mass accumulates, until at some point it reaches ~4000 K and the hydrogen starts to become ionized. At this point, the viscosity increases substantially, allowing much more efficient outwards transfer of angular momentum and inwards transfer of mass. The higher central mass accretion rate results in a luminous central x-ray source, and— because this accretion rate now exceeds the binary mass transfer rate—the mass of the disc begins to drop. At some point, the disc cools again, the viscosity drops, along with the luminosity, and the process starts again. The intervals between these cycles can last from months to centuries, depending on the mass transfer rate and binary parameters, implying that we have as yet only seen the tip of the iceberg of the transient XRB population.

Although the detailed patterns of individual outbursts (including in some cases multiple out- bursts from the same source) differ, one of the major steps of the past decade was the realization that the overall evolution of an outburst out- lined below is applicable to essentially all BHXRB outbursts (Fig. 2). Much of this global understand- ing arose from studies of one particularly bright and variable BHXRB, GRS 1915+105 (7), and our realization that what we had learned from it could be applied to other systems (8, 9). In the following sections, we describe the evolution of the outburst through Fig. 2, top; a sketch of likely geometries in the soft, intermediate/radio flaring and hard states is presented in Fig. 2, bottom. An animation of an outburst in the hardness-intensity diagram (HID) is presented in the supplementary materials (movie S1).

*The rising phase of the outburst (A → B).* Sources in quiescence are rarely regularly monitored, and so usually the first thing we know about an outburst is an x-ray source rising rapidly in luminosity, as detected by x-ray all-sky monitors. It is by now well established that BHXRBs be- low ~1% of their Eddington luminosity are al- ways in a hard x-ray state, and so the first, rising phase of an outburst takes place in this state. In the hard state, the x-ray spectrum is characterized by a spectral component that peaks in power at ~100 keV, probably as a result of inverse Compton scattering of lower-energy seed photons in a hot corona close to the black hole. The exact geometry of this corona is not clear but may correspond to a vertically extended but optically thin flow. This state shows strong and rapid variability in x-rays, with up to 50% variability on time scales between 10 ms and 100 s.

In the brightest hard states, a more blackbody-like spectral component can also be seen in very soft (<1 keV) x-rays and—sometimes— the ultraviolet, probably corresponding to a geometrically thin, optically thick accretion disc (the converse of the hot coronal component, and a good candidate to be the origin of the soft photons that are inverse Compton scattered by the corona). X-ray spectroscopy also often reveals strong iron emission lines (which can be fluorescence lines from neutral or ionized iron) in this phase. This iron line can often be fit by a relativistically broadened model, implying an origin very close to the black hole, and can in turn be used to estimate the spin of the black hole. This is because of the innermost stable circular orbit (ISCO): Within this radius, matter can no longer follow a circular orbit and will cross the black hole event horizon on very short time scales (milliseconds for a black hole of a few solar masses). The size of the ISCO depends on the spin of the black hole, ranging from 6 $R_G$ (10) for a non- rotating (Schwarzschild) black hole to 1 $R_G$ for a maximally rotating (maximal Kerr) black hole. Accurate measurements of the degree of gravitational redshift affecting the line can be used to infer how close the line is to the black hole, and from this the spin of the black hole itself, although both observation and modeling are complex. During the hard state, characteristic time scales of variability, called quasi-periodic oscillations (QPOs), are also seen to decrease, which may correspond to changing viscosity or decreasing characteristic radii in an evolving accretion disc.

In this state, sources are always observed to also show relatively steady radio emission at gigahertz radio frequencies (11). This radio emission ($L_R$) correlates in strength with the x-ray emission ($L_X$) in a nonlinear way: $L_X \propto L_R^b$, where $0.6 < b < 0.7$. In recent years, it has become apparent that a less radio-loud branch also exists in the hard state, which may have a steeper correlation (12), and yet which to date has revealed no other difference with the more radio-loud majority. The flat-spectrum synchrotron emission from the jet also appears to extend to the near- infrared (~2 mm) band.

*The hard–to-soft spectral transition (B → C → D).* At luminosities in the range of 10 to 50% of the Eddington luminosity, the x-ray spectra of BHXRBs in outburst begin to change (13). The hard x-ray component steepens and drops in luminosity as the blackbody-like component attributed to the accretion disc brightens and comes to dominate, resulting in a softening of the x-ray spectrum. Even more striking than the spectral evolution is the change in the x-ray variability properties of the BHXRB. The characteristic frequencies of variability continue to rise during the initial phases of the state transition, but at a certain point, much of the broadband noise drops away to be replaced by a single QPO, indicating strong oscillations in a relatively narrow range of frequencies (a few hertz).

During this phase, the behavior of the jet, revealed by infrared and radio observations, also begins to change. The infrared emission drops almost as soon as the state transition begins (14), indicating a change in the jet properties (density and magnetic field) close to the black hole.

The radio emission begins to vary more dramatically, showing oscillations and flare events superposed on an overall decline (8, 15). At a certain point, there are one or more large radio flares, which can be two or more orders of magnitude more luminous than the previous existing, steadier jet in the hard state. In several notable cases, high-resolution radio observations after such flares have directly resolved radio- or even x-ray–emitting blobs moving away from the central black hole (16, 17), which can be kinematically traced back to the time of the state transition. It has been recently shown that in some cases, the ejection is coincident in time with the appearance of the strong QPOs (15).

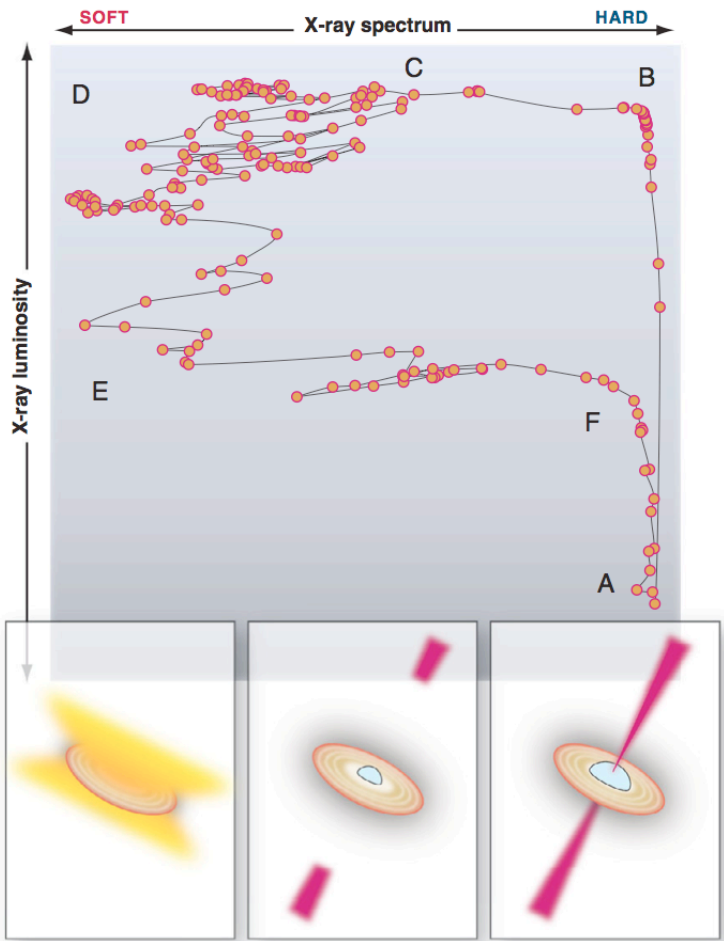

Fig. 2. The HID. The horizontal axis represents the "hardness" or color of the x-ray emission from the system, which is a crude but effective measure of the x-ray spectrum. The vertical axis represents the x-ray luminosity. BY analogy with the Hertzsprung-Russell diagram, in which the lifetime of a star can be tracked, in the HID the evolution of a black hole outburst can be tracked. Each point corresponds to a single observation. Observations on the right hand side of the HID are considered to be in "hard" x-ray states, and those on the left are in "soft" states. Although the detailed patterns of individual outbursts (including in some cases multiple outbursts from the same source) differ, one of the major steps of the past decade was the realization that the overall picture outlined below turns out to be applicable to essentially all BHXRB outbursts. (Bottom) Illustrations of likely geometries in the soft, intermediate/radio flaring, and hard states.

*The soft state (D → E).* As the spectral transition continues, these strong QPOs disappear, and the overall level of x-ray variability drops as the soft state is entered. The radio emission also fades away in most cases, probably indicating a cessation of core jet activity (18, 19).

The x-ray spectrum is now dominated by the accretion disc component, although there is a weak (a few percent of the total luminosity) high-energy tail (probably of nonthermal origin) to the spectrum that extends to megaelectron volt energies. The soft state is generally the longest lasting phase of an outburst, often persisting for a weeks at a more or less constant luminosity before beginning a slow decline. This marks the point at which the accretion rate onto the central accretor starts to drop. In the hardness-luminosity diagram (HLD) the source still makes some rapid changes in x-ray color, including occasional rapid excursions to hard x-ray states (20) that can be associated with brief reactivation of the jet as observed in the radio band (19). During the luminosity decline in the soft state, measurements of the luminosity (L) and temperature (T) of the accretion disc component follow a relation close to $L \propto T^4$, as expected for a blackbody of constantly emitting area. This is interpreted as indicating that the accretion disc has a constant radius and structure, which may correspond to the ISCO and hence can be used—as with the iron emission line in bright hard states—to estimate the black hole's spin (21).

The latest development in understanding the overall geometry and energy budget of the different states has been the realization that these soft states also ubiquitously produce a strong accretion disc wind that is focused in the plane of the accretion disc and may carry away a large amount of mass from the accretion flow (22, 23). Previously, kinetic outflow was only well established in the hard state, which shows the strong radio jet, but we now know that both states have outflows, albeit of rather different natures.

*The return to the hard state and quiescence ($E \rightarrow F$).* Eventually, as the central mass accretion rate continues to fall the BHXRB makes a transition back from the soft to the hard state. In nearly all cases, this soft $\rightarrow$ hard transition occurs at a lower luminosity than that of the earlier hard $\rightarrow$ soft transition. Furthermore, the soft $\rightarrow$ hard transition, although also showing a range of luminosities at which it can occur (even in the same source), generally occurs at a luminosity of a few percent of the Eddington luminosity (24). In fact, the soft state has never been convincingly observed in any BHXRB at luminosities below 1% Eddington. By the time the source reaches the canonical hard state again, with almost exactly the same spectral and timing characteristics as the initial hard state, the jet has reappeared, and the accretion disc wind is gone. Once in the hard state, the source decline continues, typically below the detection levels of all-sky or regular x-ray monitoring, and are observed only occasionally until their next outburst. These quiet phases are not without interest, however, for it is during these periods that— without the glare of the bright accretion disc—researchers are able to accurately measure the orbital motions of the companion star using scopes and hence estimate the mass of the black hole itself (25, 26).

These cycles of activity reveal clear changes in the way that an accreting black hole responds to an increase in accretion rate, varying the form and degree of both the radiative and kinetic (jets and winds) outputs from the liberated gravitational potential energy. These patterns have been observed more than 30 times with very few exceptions, and no system that strongly contradicts the established empirical pattern of behavior has ever been observed. These studies provide us with hope of being able to estimate the radiative and kinetic output at any given phase of, and cumulatively for, the outburst of a BHXRB. We may then use that to understand how such feed- back affects the local ambient media, energizing and accelerating particles and seeding magnetic field. We can seek to calculate how the cumulative loss of angular momentum in winds and jets may affect the evolution of the binaries. If we can truly tie properties of the accretion flow to the black hole spin, we may learn about the evolution of black holes and understand how and when to make observations to test GR in the strong field limit.

The importance of these results would be even greater if it could be demonstrated that they also shed light on accretion and feedback in the most massive black holes, those in active galactic nuclei (AGNs), which may have helped to stall cooling flows in galaxy clusters, shaping the growth of galaxies—still one of the biggest problems in extragalactic astrophysics (27). There are suggestions that this may well in fact be the case. Comparison of x-ray, optical, and radio surveys of AGNs may show similar patterns of coupling between accretion and jets (28). More quantitatively, although not without its sceptics, it has been shown that all accreting black holes fol- low an empirical relation between their mass and x-ray and radio luminosities (29, 30).

**Ultra-Luminous X-Ray Sources**

Modern x-ray observatories are easily capable of detecting the brightest x-ray sources in external galaxies. Of these, we often observe a high luminosity tail of the population at or above the Eddington limit for a stellar-mass black hole. If these ULXs contain stellar-mass black holes, then in order not to violate the Eddington limit their emission must be highly nonisotropic (beamed), or the most luminous examples must contain a black hole of much higher mass, 100 to 1000 times the mass of the Sun. A third option is that these objects do indeed emit well above the Eddington limit, which naturally would pose theoretical problems. It is most probable that the observed sample is not homogeneous and instead composed of more than one class (31). For the few most luminous, called hyper-luminous x-ray sources (HLXs), the interpretation in terms of a stellar-mass black hole appear to be difficult to reconcile with the observations. The most luminous HLX, in the galaxy ESO 243-49 (32), emits a factor of 1000 higher than the Eddington limit and shows discrete radio flaring events (33), leaving the option of a 104 $M_\odot$ black hole emitting close to the Eddington limit as the most probable. For fainter ULXs, the discussion is still open.

A number of ULXs are associated with emission-line nebulae. Because these nebulae are ionized by the radiation from the ULX, they can be used as a bolometer to estimate its total luminosity (34). From such studies, it emerges that their radiation is at most mildly beamed. Other observational evidence, such as observations of super-Eddington episodes from sources in the Milky Way and the discovery of a population of ULX associated with star-forming regions, have led to the suggestion that ULXs are a phase of the evolution of high-mass x-ray binaries. In the absence of a direct mass measurement for the compact object such as those obtained from optical observations of galactic systems, which are difficult to obtain with current optical telescopes, much of the observational effort is aimed at finding features that can be scaled to BHXRBs in our own galaxy. Different states and state transitions have been reported for several ULXs (35).

The measurement of accretion disk parameters (36) are one way to obtain an indication of the mass of the object through the inner radius of the accretion disk, which, if located at the ISCO, scales linearly with the mass of the black hole. The details of the spectral model are very complex, but very large differences can be trusted. On the side of time variability, QPOs have been detected in a few cases, most notably of the bright system X-1 in the galaxy M 82. In order to scale the observed frequencies (in the 10 mHz range) to those from bona fide stellar-mass objects, other parameters must be determined, such as the type of QPO and the association to spectral features, all of which makes this method still inconclusive.

The amount of observational data on ULXs and HLXs is still much less rich than that of stellar-mass systems, and extensive observational campaigns are needed in order to reach a better understanding of these objects and their nature.

**Conclusions and Open Issues**

*Observational perspective.* The revolution in our understanding of black hole accretion that has taken place over the past decade has been due in large part to the extremely good coverage of black hole binary outbursts by x-ray observatories. We single out the Rossi X-Ray Timing Explorer (RXTE), whose flexible scheduling and high time-resolution data led to the model presented in Fig. 2. What was clearly missing were comparably high-cadence radio and infra- red observations to track the jet as well as the accretion flow. RXTE was decommissioned in January 2012, just at the time when new radio arrays— with improved sensitivities and fields of view and whose key science goals embraced the astrophysics of variable and transient sources— were being constructed or commissioned. In the coming decade, the Square Kilometer Array precursors MeerKAT and Australian Square Kilometre Array Pathfinder (ASKAP) will provide orders- of-magnitude more radio coverage than previously has been possible (37), but RXTE will not be there to provide the x-ray data, nor is there a clear, publicly orientated replacement in the works. It can only be hoped that a new mission, with com- parable timing and all-sky monitoring capabilities, will be available in the coming decade to work with these new radio arrays.

*Key unanswered questions.* The study of black hole accretion allows us to probe both the details of GR in the strong field regime (38) and to understand the role of black holes in the conversion of gravitational potential energy into kinetic energy and radiation. This feedback of energy from these cosmic batteries is important on all scales: from heating the local interstellar medium in our own galaxy, to affecting the growth of the largest galaxies and the heating of cooling flows in the centres of galaxy clusters. To this end, understanding in more detail the relativistic jet—how it forms, and how much power it carries—is a key question. It has long been suggested that relativistic jets could be powered not by the accretion flow but by the spin of the black hole (which can contain a vast amount of energy). X-ray spectral observations can al- low us to estimate the black hole spin, and we may compare this with estimates of the jet power. It is currently unclear where this leads us: One study (39) found no correlation between jet power and reported spin measurements, whereas a subsequent study (40) has reported a correlation between transient jets and spin measurements derived from accretion disc modeling in the soft state. A key source in this story turns out to be, once again, Cygnus X-1: Recently, both disc and iron line modeling have converged on a very high spin for the black hole in this system (41, 42), which is also nearby with a very accurately measured distance (43) and has a well-studied and powerful jet (44), although as yet no major ejection events.

A long standing question for black hole accretion that remains unanswered is how much of the accreting matter and gravitational potential energy actually crosses an event horizon. Models in which liberated potential energy may be trapped in the flow and unable to escape before it crosses the event horizon appear to be able to explain why some black holes are so faint when in quiescence (45), although it turns out that such systems are also almost universally associated with a powerful jet, which could be an alternative sink for "missing" radiative output (46). The accretion disc wind in soft states also throws up some interesting questions. Is the wind responsible for halting the jet during the state transition, or are they both symptoms of something deeper? Does the existence of kinetic outflow in both soft and hard states demonstrate that such outflows, carrying away mass, energy, and angular momentum, are a necessary component for accretion? These jets should be considered in the broader context of accreting objects, such as young stellar objects, cataclysmic variables, gamma ray bursts (GRBs), unusual super- novae and—most closely related—neutron star XRBs, all of which also show jets and winds. Of these, we might hope that in the future we can link long-duration GRBs—which may correspond to a short phase of accretion at very high rates onto a newly created black hole—to the empirical models presented here. The current hot topic of possible tidal disruption events and their consequent rapid-change burst of accretion onto a supermassive black hole, with associated jet formation (47, 48), may also be possible to integrate into this broader model. Much remains to be done.

**References and Notes:**
1. Charge is assumed to be unimportant on macroscopic scales in astrophysics because the Coulomb force is much larger than the gravitational force, causing any charge separation to be rapidly neutralized.
2. S. L. Shapiro, S. A. Teukolsky, Black Holes, White Dwarfs and Neutron Stars (Wiley, New York, 1983).
3. Because the size of the black hole event horizon varies linearly with mass, the ratio of mass to radius, and hence accretion efficiency, is the same (ignoring spin for now) for black holes of all masses.
4. J. E. McClintock, R. A. Remillard, in Compact stellar X-ray sources, W. Lewin, Michiel van der Klis, Eds. (Cambridge Astrophysics Series, no. 39, Cambridge Univ. Press, Cambridge, UK, 2009) pp. 157–213.
5. OB-type stars are the hottest and most luminous, with the shortest lives.
6. The Eddington luminosity is that luminosity at which, under the simplifying assumptions of pure hydrogen accretion and spherical symmetry, the outwards radiation force (on the electrons) balances the inwards gravitational force (on the protons, to which the electrons are coupled). The Eddington luminosity scales linearly with mass, and is approximately $1.4 \times 10^{38}$ erg/s per solar mass of accretor.
7. R. P. Fender, T. M. Belloni, Annu. Rev. Astron. Astrophys. 42, 317 (2004).
8. R. P. Fender, T. M. Belloni, E. Gallo, Mon. Not. R. Astron. Soc. 355, 1105 (2004).
9. T. M. Belloni, in Lecture Notes in Physics: The Jet Paradigm (Springer, Berlin, 2010), pp. 53–81.
10. $R_G$ is the gravitational radius, $R_G = GM/c^2$ (where G is the gravitational constant, M is the mass of the black hole, c is the speed of light, and R is the radius of the black hole). It corresponds to ~1.5 km per solar mass of black hole.
11. R. P. Fender, Mon. Not. R. Astron. Soc. 322, 31 (2001).
12. E. Gallo, B. P. Miller, R. Fender, Mon. Not. R. Astron. Soc., http://arxiv.org/abs/1203.4263 (2012).
13. This transition luminosity is not fixed for the different outbursts, even for the same system (18).